# Tele-LLM-Hub: Building Context-Aware Multi-Agent LLM Systems for Telecom Networks

**Pranshav Gajjar**
NextG Wireless Lab, North Carolina State University

**Cong Shen**
Dept. of ECE, University of Virginia, WiSights Lab

**Vijay K. Shah**
NextG Wireless Lab, North Carolina State University, and WiSights Lab

## Abstract

This paper introduces Tele-LLM-Hub, a user friendly low-code solution for rapid prototyping and deployment of context aware multi-agent (MA) Large Language Model (LLM) systems tailored for 5G and beyond. As telecom wireless networks become increasingly complex, intelligent LLM applications must share a domain-specific understanding of network state. We propose TeleMCP, the Telecom Model Context Protocol, to enable structured and context-rich communication between agents in telecom environments. Tele-LLM-Hub actualizes TeleMCP through a low-code interface that supports agent creation, workflow composition, and interaction with software stacks such as srsRAN. Key components include a direct chat interface, a repository of pre-built systems, an Agent Maker leveraging fine-tuning with our RANSTRUCT framework, and an MA-Maker for composing MA workflows. The goal of Tele-LLM-Hub is to democratize the design of context-aware MA systems and accelerate innovation in next-generation wireless networks.

## 1 Introduction

The increasing complexity of modern 5G and emerging 6G networks, characterized by virtualization, disaggregation, network slicing, and dense deployments, places enormous pressure on automation. LLM-based systems have shown promise for telecom applications, but their deployment requires domain-specific grounding. Prior work, such as ORANBench13K, has demonstrated that general-purpose LLMs struggle to reason over O RAN specifications without specialized data and benchmarks Gajjar and Shah [2024]. Building on this, ORANSight introduced a suite of 18 foundational models that were trained using the RANSTRUCT framework to provide SOTA O-RAN and code generation performance Gajjar and Shah [2025]. Together, these works highlight that LLMs must be adapted to telecom data, structured specifications, and RAN semantics to be effective.

While these contributions enable powerful single-agent systems, the next frontier is coordination. Tasks such as end-to-end resource optimization, slice management, testing, anomaly detection, and security monitoring demand agents that can communicate structured knowledge among themselves. A single LLM agent is insufficient because real networks are inherently multi-faceted: one agent may reason over quality of service requirements, another over physical layer measurements, and a third over security contexts. The ability to exchange telecom-specific state information across agents is therefore fundamental. The paper Ganiyu et al. [2025] showcased how a specification-aware multi-agent LLM framework for testing and validation of O-RAN components, underscoring how coordinated LLM agents (Gen-LLM, Val-LLM, Debug-LLM) can accelerate verification of telecom standards. This observation aligns with recent surveys that emphasize communication-centric coordination as a critical capability in LLM-based MA systems Yan et al. [2025].

The telecom research community has begun exploring this direction. Wu et al. introduced LLM xApp for RAN resource management, demonstrating the feasibility of utilizing LLMs within the O-RAN ecosystem Wu et al. [2025]. Bao et al. proposed LLM-guided hierarchical RIC control, highlighting the importance of structuring interactions between near RT and non RT controllers Bao et al. [2025]. Moore et al. developed an integrated intrusion detection approach with LLM-based secure slicing xApps Moore et al. [2025], while Tang et al. designed an edge AI service provisioning framework for 6G O RAN that leverages distributed intelligence Tang et al. [2025]. These efforts collectively underline the need for structured protocols that allow LLMs and MA systems to interoperate within realistic telecom environments.

Industry voices echo this vision. Ericsson emphasizes that future autonomous networks will require conversational AI systems capable of contextual reasoning across multiple network domains Ericsson [2025], while NVIDIA blogs demonstrate telco-specific LLM applications ranging from RAG-enhanced O-RAN document processing to AI-driven network configuration NVIDIA [2024, 2025]. Collectively, both academia and industry identify the need for coordinated and context-aware LLM ecosystems in telecom.

## 2 TeleMCP

The standard Model Context Protocol (MCP) provides a strong baseline for enabling structured communication between agents in multi-agent systems. MCP establishes a common framework for agents to exchange context and coordinate effectively across diverse applications. In telecom-inspired environments, however, agents must often reason over logs, KPIs, performance counters, and user-defined metrics. To address this, we introduce **TeleMCP**, an extension of MCP that generalizes context sharing by allowing users to define which key performance indicators, parameters, and state information are important for their workflow. TeleMCP enables flexible integration of both structured inputs like a set of KPIs and unstructured data sources (e.g., logs, traces, and telemetry) from prominent RAN development stacks. These inputs are automatically ingested by the platform and exposed as shareable context objects. Using a drag-and-drop interface, users can specify which information should be exchanged between agents, ensuring transparency and control over agent communication.

## 3 Tele-LLM-Hub

We designed **Tele-LLM-Hub** as the first low-code platform that unifies agent fine-tuning, orchestration, and deployment in a drag-and-drop environment. The platform is powered by four central components. First, the Agent Maker provides a user interface for creating specialized agents. It leverages an extension of the RANSTRUCT framework, combining reasoning, instruction tuning, PEFT, and retrieval-augmented learning over a user-defined problem. The result is an agent grounded in domain knowledge and ready for integration. Second, the MA Maker is the orchestration environment where users compose workflows. Here, data sources such as network logs, srsRAN, OpenAirInterface (OAI), or external telemetry feeds are connected with logic components, including LLM agents, retrieval modules, custom analysis code, and TeleMCP nodes for structured communication. Crucially, the platform's drag-and-drop interface allows users to select the KPIs, logs, and state variables that should be shared between agents, giving fine-grained control over multi-agent interactions without requiring code. Third, the platform provides a direct chat interface for debugging and evaluating agents individually. Finally, Tele-LLM-Hub would also host a repository of pre-built systems, an exemplary implementation of one such system, *AI5GTest*, reconstructed within the MA-Maker canvas, is provided in Appendix A.

## 4 Conclusion

We introduced TeleMCP as a structured communication protocol for telecom agents and presented Tele-LLM-Hub as the first drag-and-drop low-code solution for building context-aware multi-agent LLM systems in telecom. By unifying fine-tuning, orchestration, and deployment, and by grounding agent interactions in a domain-specific context, Tele-LLM-Hub lowers the barrier for developing intelligent and collaborative applications for 5G and beyond. Future work will extend platform compatibility with more stacks and hardware, enrich the library of pre-built systems, and explore standardization of TeleMCP as a foundation for multi-agent coordination in telecom.

## A    Implementing *AI5GTest* through MA-Maker

Section 3 introduced Tele-LLM-Hub and its MA-Maker canvas for composing multi-agent workflows. This appendix provides a concrete realization of those ideas by showing how *AI5GTest* is implemented end-to-end inside MA-Maker. Figure 1 depicts the complete workflow; each block on the canvas corresponds to a configurable node whose behavior is described below.

**Input Nodes.** Input nodes accept raw artifacts. Users may choose from a wide variety of input sources and bind them to the canvas. This selection is intentionally simple; the platform surfaces the raw inputs, and the user selects which files or streams to attach to a system.

**Agent Nodes.** Agent nodes let the user add LLM agents available on the platform and configure their common attributes: system prompt, base model selection, temperature, max tokens, and decoding controls. In the AI5GTest mapping, the principal agents are Gen-LLM (generates the ordered procedural flow from the test intent), Val-LLM (validates each expected step against a selected log window), and Debug-LLM (performs deeper triangulation on failures). Each agent node stores its prompt and decoding settings with the canvas, so runs are reproducible.

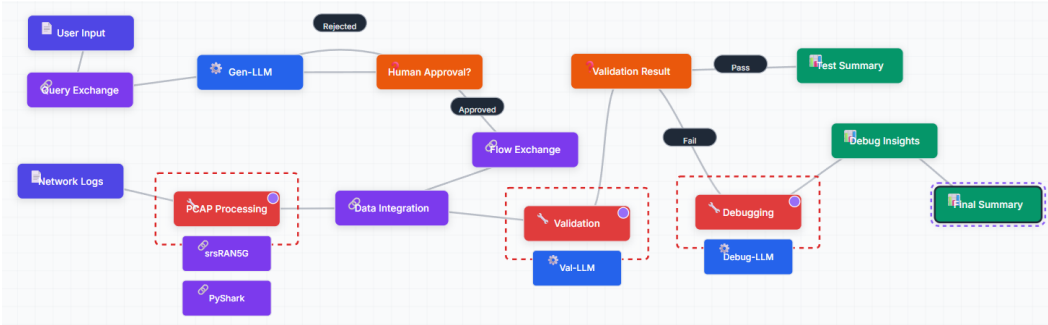

Figure 1: *AI5GTest* reconstructed within MA-Maker.

**TeleMCP Nodes.** TeleMCP nodes convert whatever raw input they receive into compact, typed context objects that downstream agents can consume. Concretely, a TeleMCP node normalizes parsed PCAP output into canonical objects such as procedural-flow (ordered step list), log-window (indexed slice of parsed messages), and message-record (protocol, name, timestamp, direction). TeleMCP thus provides a simple schema and provenance tags so the user controls what becomes shared context; users pick which TeleMCP fields to publish into the workflow rather than streaming full raw blobs to every agent.

**Logic Nodes.** Logic nodes house deterministic processing, orchestration, and small custom code. They can contain agent nodes, input references, or TeleMCP objects and are the place to express concise, bounded logic like a short Python snippet when needed. The validation loop from AI5GTest is implemented as such a logic node: the node advances a sliding window index, serializes the chosen window as a TeleMCP log window, invokes Val-LLM with the current expected step, and collects the return triplet (found/not-found, explanation, confidence). The PCAP Processing block can also be modeled as a logic node that includes two TeleMCP nodes: PyShark and srsRAN5G. These nodes are simple mapping components; they take the selected raw PCAP or trace, run custom Python code, and emit structured log indices that can be assessed by the Validation node. The user configures which parsed fields to expose (for example, message name, timestamp, and direction), and the logic node produces the indexed log-windows that are further processed by the MA system. This keeps PCAP work local to the logic node while producing standardized TeleMCP outputs that Val-LLM and Debug-LLM expect.

**Conditional Nodes.** Conditional nodes implement the decision and human-in-the-loop points. After Gen-LLM produces a candidate flow, the Human Approval conditional exposes the flow and any retrieval evidence to the user; a boolean approval TeleMCP flag advances the canvas to validation, while rejection returns a rework artifact. Conditional nodes also branch the Validation Result, *Pass* routes to reporting, *Fail* routes to debugging, and Partial can route to a configured remediation path.

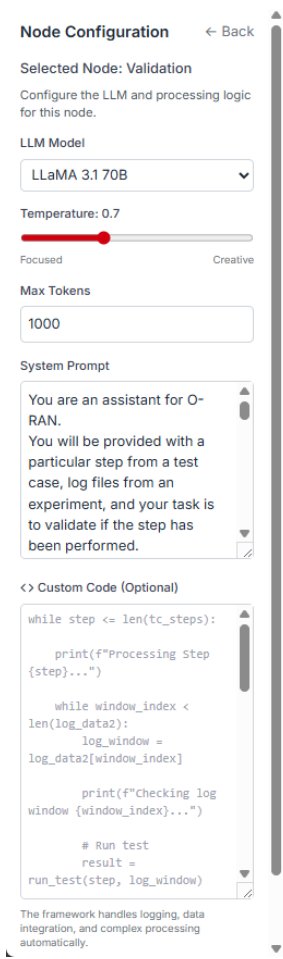

Figure 2: Configuring a logical node (Validation) in the MA-maker.

**Output Nodes.** Output nodes log and surface agent outputs. All agent outputs and node-produced artifacts are logged with provenance, so runs are auditable and exportable.

4